\documentclass[12pt,preprint]{aastex}
\usepackage{amsmath}
\usepackage{graphicx}
\usepackage{ulem}
\begin{document}

\title{Discovery of delayed spin-up behavior following two large glitches in the Crab pulsar, and the statistics of such processes}
\author{M. Y. Ge$^{1}$, S. N. Zhang$^{1,3}$, F. J. Lu$^{1}$, T. P. Li$^{1,2,3}$,
J. P. Yuan$^{4,5}$, X. P. Zheng$^{6}$, Y. Huang$^{1}$, S. J. Zheng$^{1}$, Y. P. Chen$^{1}$, Z. Chang$^{1}$
% ...
}

\affil{$^{1}$ Key Laboratory of Particle Astrophysics, Institute of High Energy Physics, Chinese Academy of Sciences, Beijing 100049, China. Email: zhangsn@ihep.ac.cn}
\affil{$^{2}$  Department of Astronomy, Tsinghua University, Beijing 100084, China}
\affil{$^{3}$  University of Chinese Academy of Sciences, Chinese Academy of Sciences, Beijing 100049, China}
\affil{$^{4}$  Xinjiang Astronomical Observatory, Chinese Academy of Sciences, Xinjiang 830011, China}
\affil{$^{5}$  Center for Astronomical Mega-Science, Chinese Academy of Sciences, Beijing, 100012, China}
\affil{$^{6}$  Institute of Astrophysics, Central China Normal University, Hubei, China}
\affil{$^{7}$  School of Physics and Technology, Wuhan University, Hubei, China}
\affil{$^{8}$ Istanbul University, Science Faculty, Department of Astronomy and Space Sciences, Beyaz{\i}t, 34119, Istanbul, Turkey}
\affil{$^{9}$  Department of Astronomy, Beijing Normal University, Beijing 100088, China}
\affil{$^{10}$  Yunnan Observatories, Chinese Academy of Sciences, Kunming, China}
\affil{$^{11}$  National Astronomical Observatories, Chinese Academy of Sciences, Chaoyang District, Beijing, China}
\affil{$^{12}$  Department of Physics, Tsinghua University, Beijing 100084, China}
\affil{$^{13}$  Key Laboratory of Radio Astronomy, Chinese Academy of Sciences, Nanjing 210008, China}
\affil{$^{14}$  National Defense University of Science and Technology, Changsha, China}
\affil{$^{15}$  Department of Engineering Physics, Tsinghua University, Beijing 100084, China}
\begin{abstract}

Glitches correspond to sudden jumps of rotation frequency ($\nu$) and its derivative ($\dot{\nu}$) of pulsars, the origin of which remains not well understood yet, partly because the jump processes of most glitches are not well time-resolved. There are three large glitches of the Crab pulsar, detected in 1989, 1996 and 2017, which were found to have delayed spin-up processes before the normal recovery processes. Here we report two additional glitches of the Crab pulsar occurred in 2004 and 2011 for which we discovered delayed spin up processes, and present refined parameters of the largest glitch occurred in 2017. The initial rising time of the glitch is determined as $<0.48$ hour. We also carried out a statistical study of these five glitches with observed spin-up processes. The two glitches occurred in 2004 and 2011 have delayed spin-up time scales ($\tau_{1}$) of $1.7\pm0.8$\,days and $1.6\pm0.4$\,days, respectively. We find that the $\Delta{\nu}$ vs. $|\Delta{\dot\nu}|$ relation of these five glitches is similar to those with no detected delayed spin-up process, indicating that they are similar to the others in nature except that they have larger amplitudes. For these five glitches, the amplitudes of the delayed spin-up process ($|\Delta{\nu}_{\rm d1}|$) and recovery process ($\Delta{\nu}_{\rm d2}$), their time scales ($\tau_{1}$, $\tau_{2}$), and permanent changes in spin frequency ($\Delta{\nu}_{\rm p}$) and total frequency step ($\Delta{\nu}_{\rm g}$) have positive correlations. From these correlations, we suggest that the delayed spin-up processes are common for all glitches, but are too short and thus difficult to be detected for most glitches.

\end{abstract}

\keywords{glitch --- stars: neutron --- pulsars: general
--- X-rays: individual (Crab pulsar)}

\section{Introduction}

Glitches are typical events of pulsars, observed as sudden
jumps in rotational frequency ($\nu$) and spin-down rate ($\dot\nu$),
usually followed by a recovery stage, in which $\nu$ and its
derivative $\dot\nu$ recover gradually to the extrapolated values of the
pre-glitch evolution trend. The behavior of the spin frequency post
glitch could be described by polynomial components and several exponential processes (as described in equation (\ref{eq00})),
such as $\sum\Delta\nu_{\rm di}\exp(-t/\tau_{\rm i})$, where $\nu_{\rm di}$
and $\tau_{\rm i}$ are amplitude and time scale
of the $i^{\rm th}$ component. Most of these exponential processes are positive values
of $\nu_{\rm di}$ (called "normal recovery processes"), however
some negative ones are observed in the Crab pulsar\citep{Lyne1992,Wong2001,Shaw2018,Zhang2018_2}.
Here, the exponential process with negative $\nu_{\rm di}$ is called
delayed spin-up process as defined in \cite{Shaw2018}.
The delayed spin-up process may dominate the evolution of $\nu$ and $\dot\nu$ immediately
after the occurrence of a glitch, but is much more difficult to be detected, probably due to that this process has a much
shorter time scale than the recovery process \citep{McCulloch1990,Dodson2002,Palfreyman2018}.
Vela pulsar is very famous for its large glitches in which
some of them have been continuously observed, but besides
the ordinary recovery processes, only upper limits of $12.6$\,s to $2$\,minutes
have been obtained for the rising time scale of these glitches, before the recovery starts.
No delayed spin up process has been detected in Vela pulsar
\citep{McCulloch1990,Dodson2002,Palfreyman2018,Ashton et al.(2019)}. The Crab pulsar is another
important object for pulsar glitch study, from which 26 glitches have been detected so far
\citep{Espinoza2011,Wang2012,Lyne2015,Shaw2018,Shaw2018b}.
Compared to Vela pulsar, Crab pulsar has two unique features though its glitch
amplitudes are usually smaller than those of Vela pulsar. The first feature is that
its $\Delta{\nu}$ and $\Delta{\dot\nu_{p}}$ values are positively and linearly
correlated \citep{Lyne2015,Shaw2018}. The second feature is that
delayed spin-up processes have been observed in its large glitches with
time scales of $0.5-3.0$\,days, such as the glitches of 1989, 1996 and
2017 \citep{Lyne1992,Wong2001,Shaw2018,Zhang2018_2}.

Presently, there are mainly two trigger mechanisms for pulsar glitches. One is the star quake
model, in which the (outer) crystalline crust of a neutron star (NS) would break as strain
in the crust gradually accumulates due to the spin-down of the NS and finally surpasses
its maximum sustainable strain. Sudden rearrangement of the stellar moment of inertia
caused by the star quake would result in a glitch \citep{Ruderman1969}. The other
mechanism invokes neutron superfluidity in a NS, which is expected when the internal
temperature of star drops below the critical temperature for neutron pairing. The superfluid
neutrons rotate by forming quantized vortices, which can get pinned to nuclei in the outer crust.
Once the pinned vortices are released suddenly, glitches are the result of angular momentum
transfer between the inner superfluid and the outer crust \citep{Anderson1975,Alpar1984a}.
After the glitch, the superfluid vortices would move outwards because of loosing angular
momentum and subsequently be repinned to the outer crust. The superfluid vortex model
has its advantage in understanding pulsar glitches, especially for the post-glitch
recovery process \citep{Baym1969}. In addition, it should be noted that sudden
crust breaking may also trigger vortex unpinning avalanches \citep{Alpar1993}.
The delayed spin-up behaviors observed in some glitches of the Crab pulsar may not be well
explained based on a simple star quake or superfluid vortex model, since the
time-scale for crust breaking and plate motion or unpinned vortices to move radially
outward is less than a minute, which is hard to account for the presence of 2-day
delayed spin-up process \citep{Graber2018}. One possible scenario for the delayed
spin-up process might be that it is the initially induced inward motion of some vortex
lines pinned to broken crustal plates moving inward towards the rotation
axis \citep{Gugercinoglu2019}. Other possible scenarios are (1) the excess heating
due to a quake in a hot crust induces secular vortex movement \citep{Greenstein1979,Link1996},
or (2) the mutual friction strength in a strongly pinned crustal superfluid region changes due
to the propagation of the unpinned vortex front \citep{Haskell2018}.

The delayed spin-up processes in glitches thus carry rich information on how the glitches
progress and thus offer valuable probes to the inner structure of neutron
stars \citep{Haskell2014,Haskell2018}. Given the small number of known
spin-up processes, any new event of this kind will add precious knowledge
about glitches and the physics behind. The common feature for the three glitches
with delayed spin-up processes happened in 1989, 1996 and 2017 is that their
$\Delta\nu$ is large, compared to the known glitches of the Crab pulsar. We therefore selected two large glitches in 2004 and 2011,
performed detailed analyses about their timing behavior, and found that they
do contain delayed spin-up processes. We name them with G1, G2, G3, G4 and G5,
corresponding to the events in  1989, 1996, 2004, 2011, and 2017, respectively.
In order to describe different components conveniently,
the full spin evolution could be divided into four components: the rapid initial spin-up process of the frequency (C1),
the delayed spin-up process (C2), exponential decay processes (C3)
and the permanent change of the frequency and its
derivatives (C4) \cite{Lyne1992,Wong2001,Shaw2018,Zhang2018_2},
which dominate the glitch behavior in different stages accordingly.
Based on the parameters of these five glitches, we have also carried out a statistical study of the
spin-up processes.

This paper is organized as follows. The observations and data reduction are described in Section 2,
and the timing analysis results are presented in Section 3. Section 4 includes the physical implications
of these results and the main conclusions.

\section{Observations and Timing Analysis}

The temporal analyses of these glitches use all the radio, X-ray and Gamma-ray  observations we can access.
We use the {\sl RXTE}, {\sl INTEGRAL} and the Nanshan 25-m radio telescope
observations, together with the spin frequency and its derivative from radio format of Jodrell Bank
\footnote{http://www.jb.man.ac.uk/~pulsar/crab.html}
\citep{Lyne1993}, to analyze the timing behaviors of G3, and due to
the low cadence, only the {\sl Fermi-LAT/GBM} observations are used
for the analyses of G4. For G5, the observations from  {\sl Insight-HXMT},
{\sl Fermi-LAT/GBM}, the Nanshan 25-m radio telescope and the
Kunming 40\,m (KM40) radio telescope are used to
study the behaviors of G5. We cite the parameters for G1 and G2 from \cite{Wong2001}.
In order to perform timing analysis, the arrival
time is corrected to Solar System Barycentre (SSB) with solar system ephemerides DE405
using the pulsar position of $\alpha=05^{h}31^{m}31^{s}.972$
and $\delta=22^{\textordmasculine}00^{\prime}52^{\prime\prime}.069$ \citep {Lyne1993}.
In this section, we first describe the data reduction for observations. Then, the calculation for time of arrival
(TOA) and its error are presented. Finally, the description of the timing method for the glitches is given.

\subsection{Data Reduction for Radio observations}

We unitize the radio observations from Nanshan 25-m radio telescope located in China \citep{Wang2001}
to supply timing solution for G3 and G5. We also utilize some observations from Kunming 40\,m (KM40)
radio telescope located in China \citep{Wang2001,Xu2018} to supply timing solution for G5.

The Nanshan 25-m radio telescope, operated by Xinjiang Astronomical
Observatory (XAO), has observed the Crab pulsar frequently since
January 2000 \citep{Wang2001}. The two hands of linear polarization are obtained with
a cryogenically cooled receiver at center frequency of 1540\,MHz with
bandwidth 320\,MHz. The signals are fed through a digital filter bank
with configuration of 2$\times$1024$\times$0.5\,MHz for pulsar timing.
The samples are 8-bit digitized at 64\,$\mu$s interval and written as PSRFITS file \citep{Hotan2004}.
The integration time of each observation of the Crab pulsar is 16 minutes.

The timing observations at 2256\,MHz were conducted with the Kunming 40\,m (KM40) radio
telescope \citep{Xu2018}, operated by Yunnan Astronomical Observatory.
A room temperature receiver provides circularly-polarized signal with bandwidth of 140\,MHz.
The digital filter band divides the intermediated frequency signal
with 1.0\,MHz for each sub-channel. The integration time of
each observation of the Crab pulsar is 48\,minutes.

For the radio observations, the off-line data reduction is performed
in the following two steps using the PSRCHIVE package \citep{Hotan2004}:
(1) the data are de-dispersed and summed to produce a total intensity profile;
(2) correlate the data with the standard pulse profiles of the Crab pulsar
to determine the local TOAs that correspond to the peak of the main pulse. The detailed data
reduction process is the same as that described in \cite{Yuan2010}.

\subsection{Data Reduction of X-ray and $\gamma$-ray observations}
In this section, we introduce the data reduction processes of the
RXTE, INTEGRAL, Insight-HXMT and Fermi observations, respectively.

\subsubsection{Data Reduction of the {\sl RXTE} Observations}

The {\sl RXTE} observations used in this paper were obtained by
both the Proportional Counter Array (PCA) and the High Energy X-ray
Timing Experiment (HEXTE). The detailed introduction of PCA and HEXTE
can be found in \cite{Rothschild et al.(1998)},
\cite{Jahoda et al.(2006)} and \cite{Yan et al.(2017)}.
In this paper, the public data (ObsID P80802 and P90802)
in event mode E\_250us\_128M\_0\_1s in 5--60\,keV from PCA and
E\_8us\_256\_DX0F in 15--250\,keV from HEXTE are used.
The Standard {\sl RXTE} data processing method with HEASOFT (ver 6.25 )
is used to obtain the timing data ( i.e., the arrival time of each photon used in the analyses) as follows:
(1) Generate the Good Time Interval by ftool \texttt{maketime}
based on the {\sl RXTE} filter file. (2) Filter the events with the \texttt{grosstimefilt} tools;
(3) Convert the arrival time of each photon to the
Solar System Barycenter (SSB) with \texttt{faxbary}.
The criteria of the selection and the detailed process can be found in \cite{Yan et al.(2017)}.
The TOA for RXTE is integrated from the typical observation.

\subsubsection{Data Reduction for INTEGRAL}
The INTEGRAL observations of the Crab pulsar are subdivided into the so-called Science
Windows (ScWs), each with a typical duration of a few kiloseconds \citep{Winkler2003}.
By selecting offset angles to the source of less than 10 degrees, between
2014-03-01 and  2014-04-01, 96 public ScWs are selected for Crab
in the data archive at the INTEGRAL Scientific Data Center.
The data reduction is performed using the standard Off-line Scientific
Analysis (osa), version 10.2. The integration time of TOA for INTEGRAL
is about 1\,hour.

\subsubsection{Data Reduction for Insight-HXMT}

Launched on June 15, 2017, {\it Insight-HXMT} was originally
proposed in the 1990s, based on the Direct Demodulation
Method \citep{Li93,Li94}. As the first X-ray astronomical
satellite of China, {\it Insight-HXMT} carries three main payloads
onboard \citep{Zhang2014,Zhang2017,Liu2019,Chen2019,Cao2019}: the High Energy X-ray telescope
(HE, 20-250\,keV, 5100\,cm$^{2}$), the Medium Energy X-ray telescope
(ME, 5-30\,keV, 952\,cm$^{2}$), and the Low Energy X-ray telescope
(LE, 1-15\,keV, 384\,cm$^{2}$). The data reduction
for the Crab observations is done with HXMTDAS software v1.0 and
the data processes are described in \cite{Chen2018}, \cite{Huang2018} and \cite{Tuo2019}.
One TOA is obtained from the typical observation.

\subsubsection{Data Reduction for Fermi-LAT/GBM}

The Large Area Telescope (LAT) is the main instrument of Fermi, which
can detect $\gamma$-rays in the energy range from 20\,MeV to 300\,GeV
and has an effective area of $\sim 8000$\,cm$^2$.
It consists of a high-resolution converter tracker, a CsI(Tl) crystal calorimeter,
and an anti-coincidence detector, which make the directional
measurement, energy measurement for $\gamma$-rays, and background
discrimination, respectively \citep{Atwood2009}.

In this work, we use the LAT data to perform timing analysis with the Fermi Science Tools (v10r0p5)
\footnote{https://fermi.gsfc.nasa.gov/ssc/data/analysis/scitools/pulsar\_analysis\_tutorial.html}.
The events are selected with the angular distance less than 1\textordmasculine
\,\,of the Crab pulsar and a zenith angle of less than 105\textordmasculine\,\, and
energy range 0.1 to 10\,GeV \citep{Abdo2010}. After event selection, the
arrival time of each event is corrected to SSB with DE405. One TOA is
obtained from every two-day exposure.

\begin{figure}
\begin{center}
\includegraphics[scale=0.5]{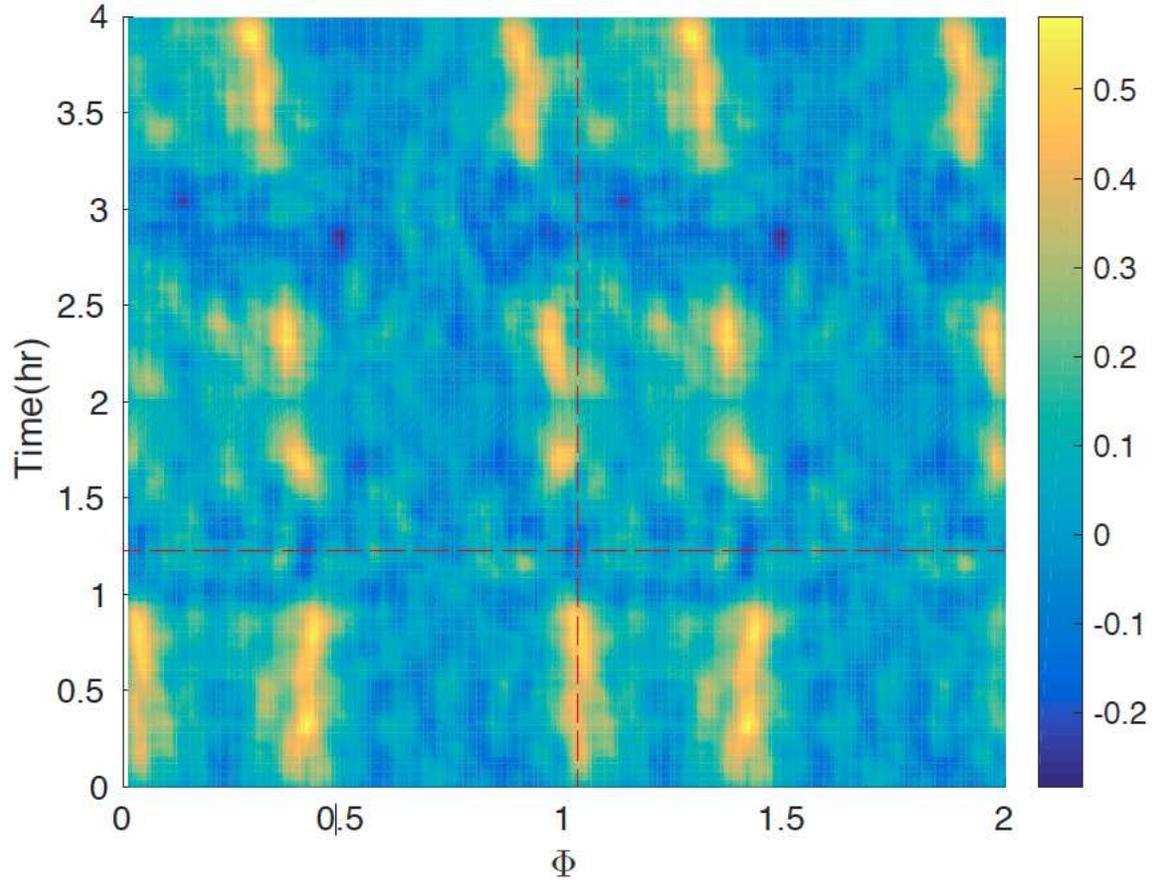}\caption{
The pulse profiles detected by Fermi/GBM as function of time around the glitch
of G5. Two periods are plotted in this figure. Each profile is integrated with 10\,minutes.
The vertical dashed line around 1.0 marks the peak position. The horizontal line
represents the glitch epoch of G5. The pulse signal around 1\,hour and 3\,hours disappears
because the Crab pulsar could not always be in the field view of Fermi satellite and
could be occulted by the Earth. This image is smoothed with gaussian
function with radius 10 pixels to eliminate the effect of fluctuation due to 1000 bins for
every profile.
\label{fig0}}
\end{center}
\end{figure}

We also utilize the Gamma-ray Burst Monitor (GBM) data
around the glitch epoch to refine the timing results.
Due to the large field of view ($\sim 2\pi$) of GBM and its
relatively high count rate ($\sim 30$\,cnts/s) of the Crab pulsar,
GBM can also be used to monitor the Crab pulsar
continuously like LAT and even has higher cadence as shown in Figure \ref{fig0}.
Given the periodicity of the pulse signals, they could be detected when the pulsar
is in the field of view of GBM, though the overall background is high due to the large field of view of GBM.
As the volume of GBM data is very large, we only select
one month data around G4 and G5, which cover 10\,days before each glitch epoch
and 20\,days after glitch epoch. The events with elevation angle greater
than 5 degrees are used to perform timing analysis. Then, one TOA can
be accumulated every 10\,minutes observation.

\subsection{TOA calculation for X-ray and $\gamma$-ray observations}
{\bf The evolutions of the spin frequency and its derivatives are estimated from
the TOAs utilizing the timing tool TEMPO2 \citep{Hobbs et al. (2006)}, while the
TOAs are obtained in a similar way to that in \cite{Ge2019}:
we first obtained a standard pulse profile that contains 100 bins from all observations,
then calculated the phase shift $\Phi_{0}$ in each observation using its pulse profile, 
the standard pulse profile and the cross correlation method, and finally the TOA is calculated 
with the formula ${\rm TOA=T_{0}+\Phi_{0}/\nu/86400}$, where ${\rm T_{0}}$ is the start time of 
one observation and ${\rm \nu}$ is the spin frequency. The uncertainty of a TOA is calculated with 
a Monte-Carlo method as also described in \cite{Ge2019}.}

\subsection{Timing Analysis}
\subsubsection{Part--Timing Analysis}
We apply the part--timing method to show the spin evolution versus time
as described in \cite{Ferdman2015}. In order to show spin evolution directly,
we divide the data set into several subsets for each glitch.
For G3, the time step for $\nu$ and $\dot\nu$ by TEMPO2 \citep{Hobbs et al. (2006)}
is about 15\,days without data over-lapping due to the low cadence of the observations.
With high cadence of the observations for G4 and G5, the time steps of the dataset for
G4--5 are chosen as 1.5 and 0.5\,days, respectively. For $\dot\nu$,
the time steps of the dataset for G4--5 are chosen as two times
as $\nu$, and the overlapping time is set as equal to the time step
in order to show more data points in the figures.
For each subset, we have taken the center of the time span as
the reference epoch for the timing analysis.

\subsubsection{Coherent Timing Analysis}

A coherent timing analysis of the data set is performed, in order to obtain more
precise measurements of the glitch parameters using TEMPO2. The phase evolution of the glitch
could be described as equation (\ref{eq00}) considering the glitch parameters \citep{Wong2001}.
\begin{equation}
\Phi=\Phi_{0} + \nu{(t-t_{0})}+\frac{1}{2}\dot\nu{(t-t_{0})^{2}}+\frac{1}{6}\ddot\nu{(t-t_{0})^{3}} + \Phi_{\rm g}(t),
\label{eq00}
\end{equation}
where $\nu$, $\dot\nu$ and $\ddot\nu$ are the spin parameters at the epoch $t_{0}$. $\Phi_{\rm g}(t)$ is the phase
description after the glitch as defined in equation (\ref{eq01}).
\begin{equation}
\Phi_{\rm g}(t)=\Delta\nu_{\rm p}\Delta{t}+\frac{1}{2}\Delta\dot\nu_{\rm p}\Delta{t}^2+
\sum_{i=0,1,2}{\Delta{\nu}_{\rm di}\tau_{\rm i}(1-\exp(-\Delta{t}/\tau_{\rm i})}),
\label{eq01}
\end{equation}
where $\Delta{t}=t-t_{\rm g}$ is the time after glitch,
$\Delta\nu_{\rm p}$ and $\Delta\dot\nu_{\rm p}$ are the permanent changes of $\nu$ and $\dot\nu$ for C4,
$\tau_{0}$, $\tau_{1}$ and $\tau_{2}$ are the time scales for C1, C2 and C3,
$\Delta{\nu}_{\rm d0}$, $\Delta{\nu}_{\rm d1}$ and $\Delta{\nu}_{\rm d2}$ are the amplitudes of the three components.
$i=1$ refers to the delayed spin-up process, and $i=2$ refers to the conventionally
observed exponential recovering process. In the following timing analysis, the effect of C1 is
neglected as its time scale is too short, which will be analyzed in Section 3.2.

In order to obtain the net evolution of a glitch, we subtract the pre-glitch
spin-down trend and then fit the frequency residuals $\delta\nu$ with equation (3) \citep{Lyne1992,Wong2001,Xie2013},
which consists of a linear function and two exponential functions,
\begin{equation}
\delta{\nu}=\Delta\nu_{\rm p}+\Delta\dot\nu_{\rm p}\Delta{t}+\sum_{i=1,2}{\Delta{\nu}_{\rm di}\exp(-\Delta{t}/\tau_{\rm i})},
\label{eq0}
\end{equation}
where the parameters have the same definition with equation (\ref{eq01}).

The coherent timing analysis for different instruments is performed simultaneously
using parameter `JUMP' to describe the time lags between different energy bands because
peak position of the Crab pulsar evolves with energy as reported in
\cite{Kuiper et al.(2003),Molkov et al.(2010),Ge et al.(2012)}. Setting the position of the
radio peak as phase 0, the values of JUMP are -0.340\,ms ( Insight-HXMT/RXTE/GBM),
-0.275\,ms ( INTEGRAL), -0.250\,ms (LAT), compared to radio band, respectively.

The residual $\delta\dot{\nu}$ can be described by
\begin{equation}
\delta\dot{\nu}=\Delta\dot\nu_{\rm p}+\sum_{\rm i=1,2}{\Delta{\dot\nu}_{\rm di}\exp(-\Delta{t}/\tau_{\rm i})},
\label{eq1}
\end{equation}
where $\Delta{\dot\nu}_{\rm di}=-\Delta{\nu}_{\rm di}/\tau_{\rm i}$.
The total frequency and frequency derivative changes at the time
of the glitch are $\Delta\nu_{\rm g}= \Delta\nu_{\rm p}+\Delta\nu_{\rm d1}+\Delta\nu_{\rm d2}$
and $\Delta\dot\nu_{\rm g}= \Delta\dot\nu_{\rm p}+\Delta\dot\nu_{\rm d1}+\Delta\dot\nu_{\rm d2}$,
respectively; and the degree of recovery can be described by
parameters: $\hat{Q}=\Delta\nu_{\rm d2}/(\Delta\nu_{\rm g}+|\Delta\nu_{\rm d1}|)$, as
suggested by \cite{Wong2001}.

\section{Results}
\subsection{The timing results for G3--5}

\begin{figure}
\begin{center}
\includegraphics[scale=0.75]{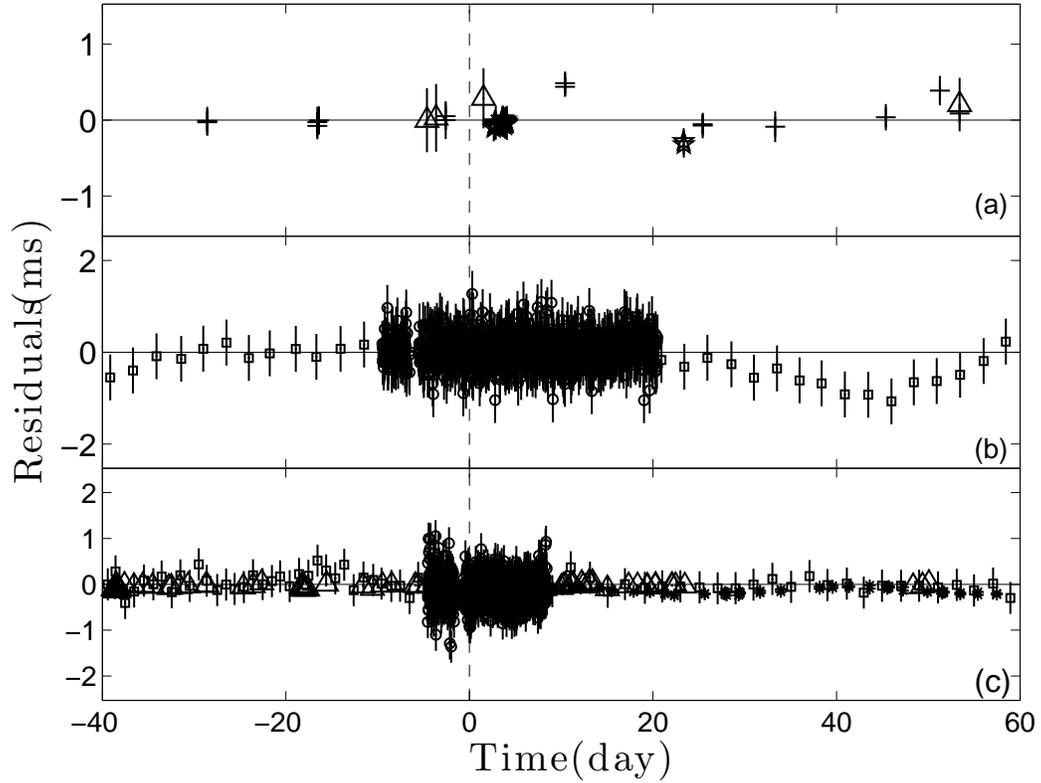}\caption{
The timing residuals. The timing residuals for G3, G4 and G5
are shown panels (a), (b) and (c). The time for glitch epochs
are set 0 to show the residuals in the same time range marked
by vertical dashed lines. The circle, square, plus, pentagram and triangle points
represent the observations from Fermi-GBM, Fermi-LAT, RXTE, INTEGRAL and radio telescopes.
\label{fig0_0}}
\end{center}
\end{figure}

\begin{figure}
\begin{center}
\includegraphics[scale=0.6]{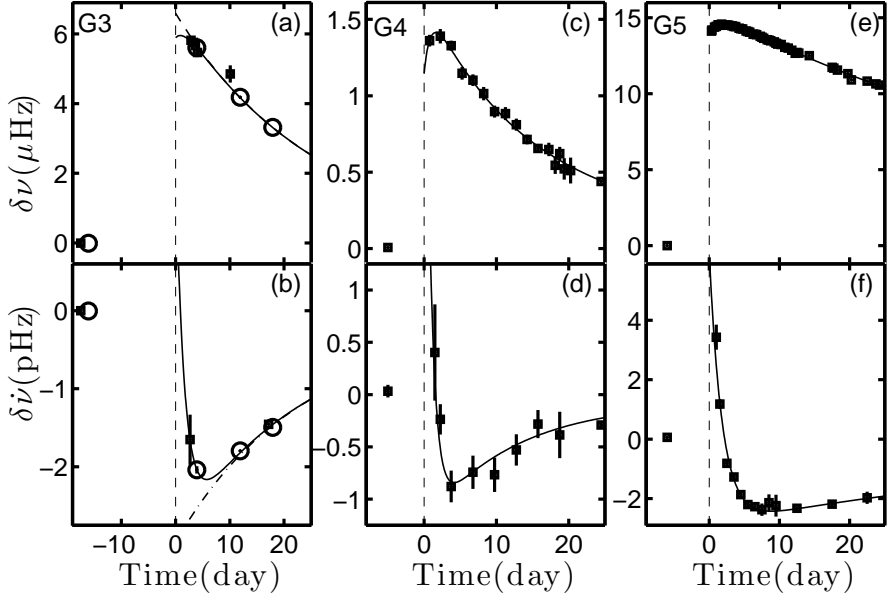}\caption{
The spin evolution of G3 (MJD 53067), G4 (MJD 55875) and G5 (MJD 58064).
Panels (a) and (b) are the evolution of spin frequency $\nu$ and frequency
derivative $\dot{\nu}$ with fitting result subtracted from the pre-glitch parameters
for G3. The square points are the part-timing results from {\sl RXTE}, {\sl INTEGRAL} and
Nanshan radio observations. $\delta{\nu}$ and $\delta{\dot\nu}$ marked by
empty circle points are the results from monthly ephemerides of Jodrell Bank (http://www.jb.man.ac.uk/pulsar/crab/crab2.txt).
For both panels, thin line represents the fitting result with equations (\ref{eq0}) and (\ref{eq1}),
respectively. The dot-dashed line represents the fitted result without spin-up process.
Panels (c)  and (d) are similar to panels (a) and (b), but for G4. The results of G4 are
obtained from {\sl Fermi} data. Panels (e)  and (f) are similar with
panels (a) and (b), but for G5. The results of G5 are
obtained from the {\sl Insight-HXMT}, Radio Telescopes in
Xinjiang and Kunming China and {\sl Fermi}. The vertical lines in all
panels represent the glitch epochs.
\label{fig1}}
\end{center}
\end{figure}

We first analyze G3, the second largest glitch, by using the coherent timing method.
The timing residuals are shown in Figure \ref{fig0_0}(a) and the timing parameters are
listed in Table \ref{table_timing_para}. The parameters of C3 are consistent with the result from \cite{Wang2012}.
After subtraction of pre-glitch evolution, the residuals $\delta{\nu}$ and $\delta{\dot\nu}$
are plotted in Figure \ref{fig1} (a) and (b). Due to the observational coverage, no
spin-up process has been detected for $\delta\nu$ and marginally for $\delta\dot{\nu}$.
However, the observational data can not be acceptably fitted without C2 with reduced $\chi^{2}$ 1.3 ( d.o.f=54),
which means that the delayed spin-up process is needed.
Fitting the data with both C2 and C3 gives the time scale $\tau_{1}$ of G3 as $1.7\pm0.8$\,days and
$\Delta\nu_{\rm d1}=-0.35\pm0.05$\,$\mu${Hz} for the delayed spin-up process.
We note here that the delayed spin-up process of G3 could be quantified in
more details, by using data such as the daily radio monitoring observation
at Jodrell Bank observatory.

G4 is also analyzed using the coherent timing method. The timing residuals are shown
in Figure \ref{fig0_0}(b) and the timing parameters are listed in Table \ref{table_timing_para}.
As shown in Figure \ref{fig1} (c) and (d) for G4, after subtraction of the pre-glitch
evolution, the spin frequency residual $\delta{\nu}$ increases first
and then decreases with time, which is just the feature of the delayed spin-up
process. From the coherent timing analysis, we can obtain that $\tau_{1}=1.6\pm0.4$\,days
with $\Delta\nu_{\rm d1}=-0.43\pm0.05$\,$\mu${Hz}. With the fitted parameters,
$\delta{\dot\nu}$ can also be described by equation (\ref{eq1}) with the
same parameters as shown in Figure \ref{fig1} (d).

G5 is re-analyzed using the coherent timing method as well. The timing residuals
are show in Figure \ref{fig0_0}(c) and the timing parameters are
listed in Table \ref{table_timing_para}.
As shown in Figure \ref{fig1} (e) and (f), the evolution of frequency residual
$\delta{\nu}$ is consistent with the result reported in \citep{Shaw2018,Zhang2018_2}.
From the fitting result,
the time scale $\tau_{1}$ for the delayed spin-up process is
$2.56\pm0.04$\,days and $\Delta\nu_{\rm d1}=-1.23\pm0.01$\,$\mu${Hz},
which are also consistent with the result of \cite{Shaw2018} and \cite{Zhang2018_2}.
The rest of parameters are listed in Table \ref{table_timing_para}.

Our analysis shows that G3 and G4 also have delayed spin-up process.
Including G1, G2 and G5, there are five glitches with delayed spin-up process.
From Table \ref{table_timing_para}, the mean time scale $\tau_{1}$ of C2 is $\sim1.4$\,days
while the mean amplitude $\Delta\nu_{\rm d1}$ of C2 is around $-0.6$\,$\mu$Hz\citep{Lyne1992,Wong2001,Shaw2018,Zhang2018_2}

\subsection{The rising time constraint of C1}
\begin{figure}
\begin{center}
\includegraphics[scale=0.6]{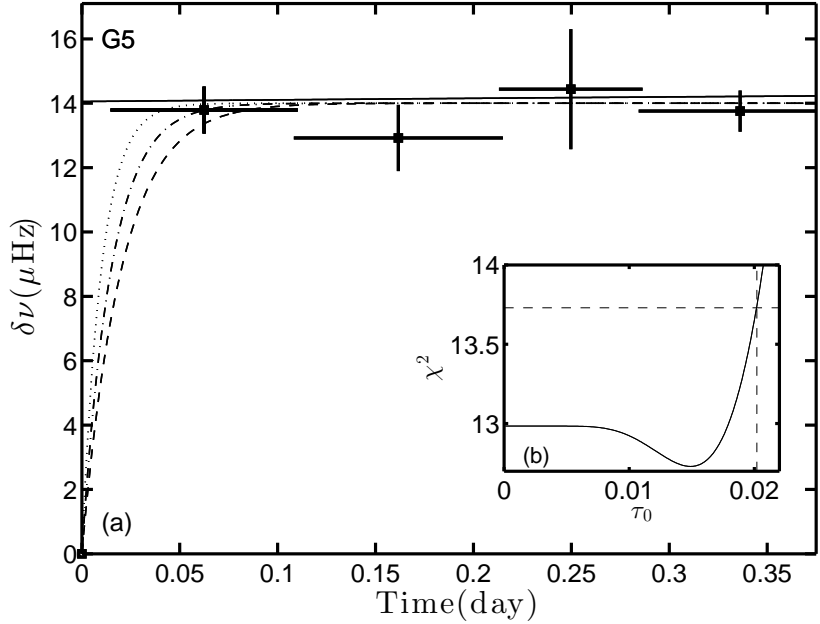}\caption{
The spin evolution of G5 just post the glitch.
Panel(a): Similar with Figure \ref{fig1}(e), $\delta{\nu}$ is the evolution of spin
frequency $\nu$ with fitting result subtracted from the pre-glitch parameters.
The thin line is spin evolution with parameters listed in Table \ref{table_timing_para}.
The dotted, dot-dashed and dashed lines represent the equation (\ref{eq2})
with time scale $\tau_{0}=0.01,0.015,0.0202$\,days, respectively. Panel (b): $\chi^{2}$
as function of $\tau_{0}$ to fit $\delta{\nu}$ with the equation (\ref{eq2}).
\label{fig2}}
\end{center}
\end{figure}

The rising time scale of C1 is very important to study the pinning process
between the inner superfluid and outer crust \citep{Haskell2018}. We make use of the observations
from {\sl Fermi-GBM} to constrain the rising time of C1 for G5. Unfortunately,
the Crab pulsar were occulted by earth at the right time for G5 in {\sl Fermi-GBM} observation.
In order to constrain the rising time scale of C1 of G5, equation (\ref{eq2}) is used
to describe the frequency evolution of C1 \citep{Haskell2018}.
\begin{equation}
\delta{\nu}=\Delta{\nu}_{0}(1-\exp(-\Delta{t}/\tau_{0})),
\label{eq2}
\end{equation}
where $\Delta{\nu}_{0}$ is the amplitude of the frequency jump and $\tau_{0}$ is the rising time scale
and $\Delta{t}=t-t_{\rm g}$ is the time after glitch. As shown in Figure \ref{fig2} (a) and (b), $\delta{\nu}$
could be fitted with equation (\ref{eq2}). As shown in Figure \ref{fig2} (b), the rising time scale
$\tau_{0}$ is less than 0.0202\,day (0.48\,hour), which is much less than the upper limit of 6 hours
given by \cite{Shaw2018} but still longer than the theoretical value of 0.1\,hour
suggested by \cite{Graber2018} and \cite{Haskell2018}.
For G3, the rising time scale could not be constrained because no high
cadence observations could be obtained in high energy bands and radio bands from
Nanshan 25-m radio telescope around the glitch epoch. For G4, the errors of $\delta{\nu}$
is close to the frequency step with short integrated time 10 minutes.

\subsection{The correlations between the parameters of C2 and other components}

We first compare the relationship between G1--5 and the other glitches of the Crab
pulsar, to see how these five glitches differ from the other ones.
The most conventional comparison is to study the jump amplitudes
of their frequencies and frequency derivatives. As shown in Figure \ref{fig3},
the Pearson coefficient between $|\Delta{\dot\nu_{\rm g}}|$ and $\Delta{\nu_{\rm g}}$ is 0.81.
Hence, $|\Delta{\dot\nu_{\rm g}}|$ and $\Delta{\nu_{\rm g}}$ show strong linear correlation
for all the glitches of the Crab pulsar, including those with delayed spin-up processes.
We also compare the correlation between $|\Delta{\dot\nu_{\rm p}}|$ and $\Delta{\nu_{\rm g}}$,
which is similar to Figure 5 in \cite{Lyne2015}. The value of $|\Delta\dot\nu_{p}|$ is obtained
from \cite{Wong2001}, \cite{Wang2012} and this work because the calculation process
in \cite{Lyne2015} is different from the rest ones. As shown in Figure \ref{fig3},
$|\Delta{\dot\nu_{\rm p}}|$ has strong linear correlation with $\Delta{\nu_{\rm g}}$ as
the Pearson coefficient is 0.98.

As shown in Figure \ref{fig3_2}, the $\Delta{\nu_{\rm g}}$ and $|\Delta{\dot\nu_{\rm g}}|$
values for the five glitches with delayed spin-up process locate in the higher wing of
the overall distribution of all the glitches and are not well separated from those
with the rest glitches. This unified positive correlation suggests that the physical
mechanism of the five glitches with delayed spin-up processes
is probably the same as that of all the other glitches, and it is
worth to check from the archival data whether the glitches
occurred in 1975, 2000, 2001 and 2006 also have delayed spin-up processes,
as they have amplitudes comparable to those of G2.

To understand more characteristics for G1--5,
we examine the Pearson and Spearman correlations between their parameters as listed in Tables
\ref{corre_para0}, \ref{corre_para1} and plotted in Figures \ref{fig4}, \ref{fig5}. The
relationships between $\tau_{1}$, $|\Delta{\nu}_{\rm d1}|$, $\tau_{2}$, $\Delta{\nu}_{\rm d2}$,
$\Delta{\nu}_{\rm p}$, $|\Delta{\dot\nu}_{\rm p}|$ and $\Delta{\nu}_{\rm g}$
have positive correlations as shown in Figures \ref{fig4} and \ref{fig5}, some
of which are consistent with the result of \cite{Wang2019}.
These positive correlations mean that C2 has a larger
amplitude and longer time scale when a glitch has a larger
spin frequency jump. If C2 also exists for the smaller
glitches, from the positive correlation between $\tau_{1}$
and $\tau_{2}$, $\tau_{1}$ should be less than 0.5\,days if $\tau_{2}<10$\,days,
which indicates that C2 might not be easily observed
due to the low cadence of most previous observations.

We also find that the correlations between $\Delta{\dot\nu}_{\rm d1}$ and
$|\Delta{\dot\nu}_{\rm d2}|$, $|\Delta{\dot\nu}_{\rm p}|$, $\tau_{1}$, $\tau_{2}$, $\hat{\rm Q}$
are weak listed in Tables \ref{corre_para0} and \ref{corre_para0}.

\section{Discussions and Summary}

It is generally believed that a neutron star has the following interior structure: the
outer crust made by degenerated electrons and an ion crystal lattice, the inner
crust composed of nucleus, superfluid neutrons, probably superfluid
protons and leptons, the outer core that contains superfluid neutrons,
superfluid protons and electrons, and the inner core \citep{Anderson1975,Alpar1984a}.
The angular momentum transfer from the inner superfluid component to the outer
normal component can explain the observed frequency jumps (glitches)
of pulsars \citep{Anderson1975,Alpar1981}. The response to the glitch of the thermal
vortex creep process in the pinned superfluids are suggested to be responsible
for the post-glitch behaviors \citep{Alpar1984a,Alpar1984b,Larson2002}.
A quick rise of the spin rate in crust, resulting from the initial energy deposition, could be followed by a
slower rise as the thermal wave dissipation in the effective crust with thickness ~200m,
depending on the crust equation of state \citep{Link1996,Larson2002}.
Another possible scenario is that vortex accumulation
in strong pinning regions leads to differential rotation and the propagation
of vortex fronts, which naturally produces a slower component of the rise
after the initial fast step in frequency jump \citep{Haskell2014,Khomenko2018,Haskell2018}.
Recently, the combination of crust-quake vortex and unpinning models is proposed to explain the
whole glitch behavior as suggested by \cite{Gugercinoglu2019}.
\cite{Haskell2018} estimated the rising time scale of rapid initial spin-up of the largest glitch G5 is
$\sim0.1$\,hours, which is consistent with upper limit of 0.48\,hours for G5.
We hope that the positive correlations between the amplitudes and time scales
of C2 and C3 can be also used to constrain the properties of neutron star structures.

\begin{figure}
\centering
\caption{{The $|\Delta{\dot\nu_{\rm g}}|$ or $|\Delta{\dot\nu_{\rm p}}|$ and $\Delta{\nu_{\rm g}}$ correlations for the Crab pulsar.}
The samples are maintained from the website
http://www.jb.man.ac.uk/pulsar/glitches.html \citep{Espinoza2011}.
Black circle points are the glitch events of the Crab pulsar and
dual-circles represent G1--5.
Black square points are the glitch events of the Crab pulsar but for $|\Delta{\dot\nu_{\rm p}}|$.
The value of $|\Delta{\dot\nu_{\rm p}}|$ is obtained from \cite{Wong2001}, \cite{Wang2012} and this work. Dual-square points represent G1--5 but for $|\Delta{\dot\nu_{\rm p}}|$. 
}
\includegraphics[width=0.6\textwidth]{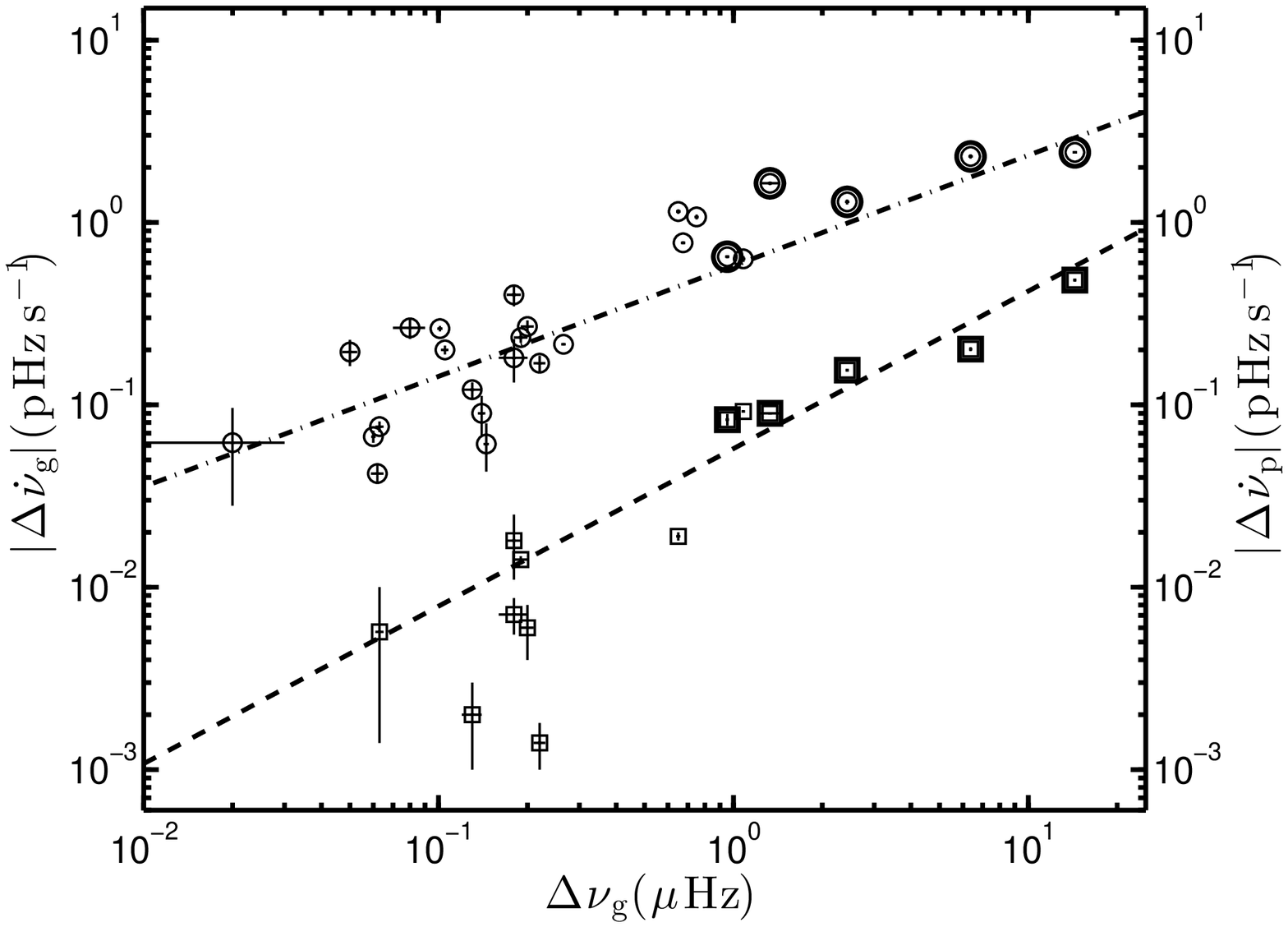}
\label{fig3}
\end{figure}

\begin{figure}
\centering
\caption{The distributions of $\Delta{\nu_{\rm g}}$ and $|\Delta{\dot\nu_{\rm g}}|$ of the Crab pulsar.
Panel (a): the solid-black, dashed-blue and dashed-red lines represent
the distribution of $\Delta{\nu_{\rm g}}$, with delayed spin-up process and the rest glitches, respectively.
Panel (b): The same description as panel (a) but for $|\Delta\dot{\nu_{\rm g}}|$.
}
\includegraphics[width=0.6\textwidth]{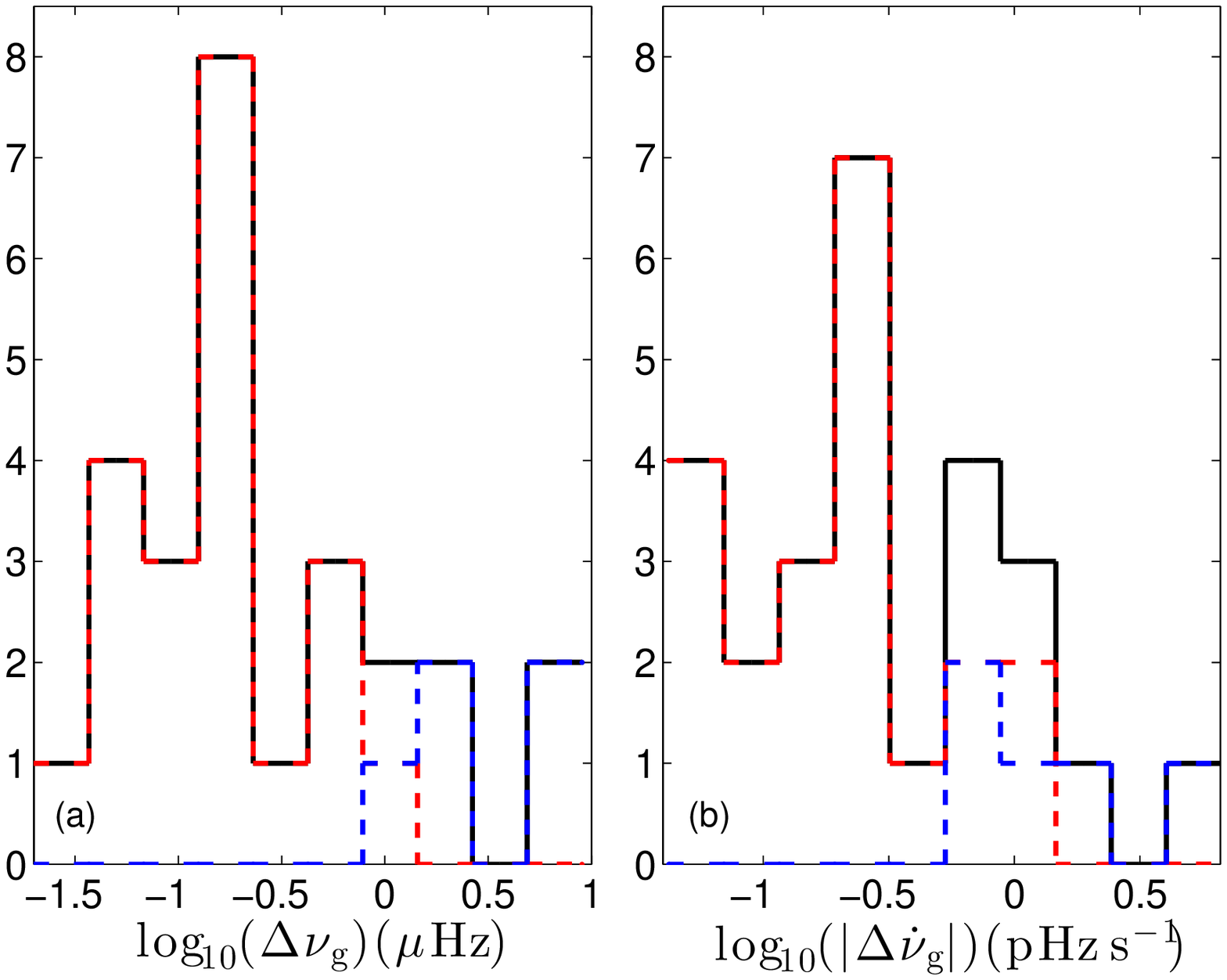}
\label{fig3_2}
\end{figure}

\begin{figure}
\begin{center}
\includegraphics[scale=0.6]{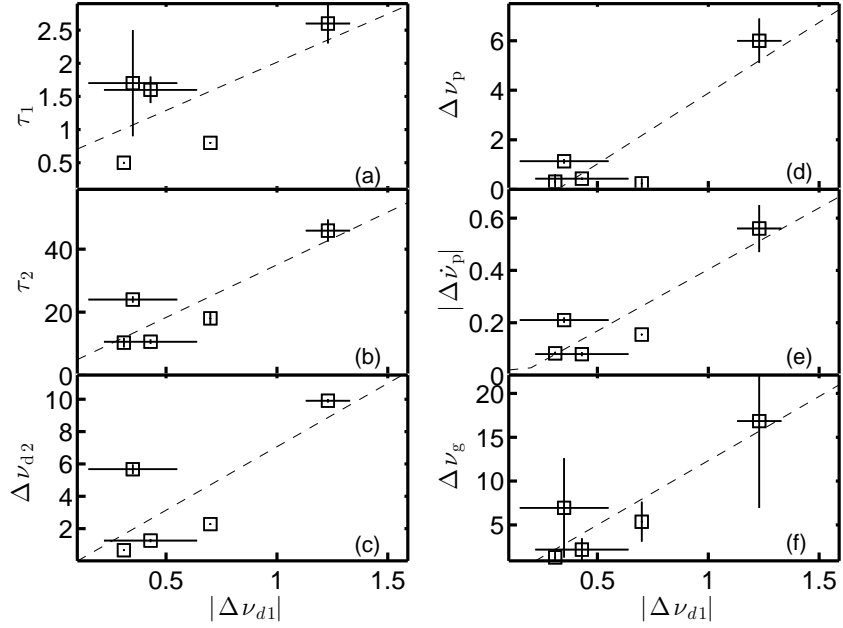}\caption{
The correlations between the parameters of G1--5 with delayed
spin-up processes. The unit of $|\Delta\nu_{\rm d1}|$, $\Delta\nu_{\rm d2}$, $\Delta\nu_{\rm p}$ and  $\Delta\nu_{\rm g}$
is $\mu$Hz; $\Delta\dot\nu_{\rm p}$ is in units of pHz\,s$^{-1}$; $\tau_{1}$ and $\tau_{2}$ are in units of day.
The dashed lines in panels (a)--(f) are the linear fitting
results for the correlations. The errors of $\tau_{1}$ and $\Delta\nu_{\rm d1}$ for
1989 (G1) and 1996 (G2) are taken as zero, since \cite{Lyne1992} and \cite{Wong2001}
did not report them in their works.
\label{fig4}}
\end{center}
\end{figure}

\begin{figure}
\begin{center}
\includegraphics[scale=0.6]{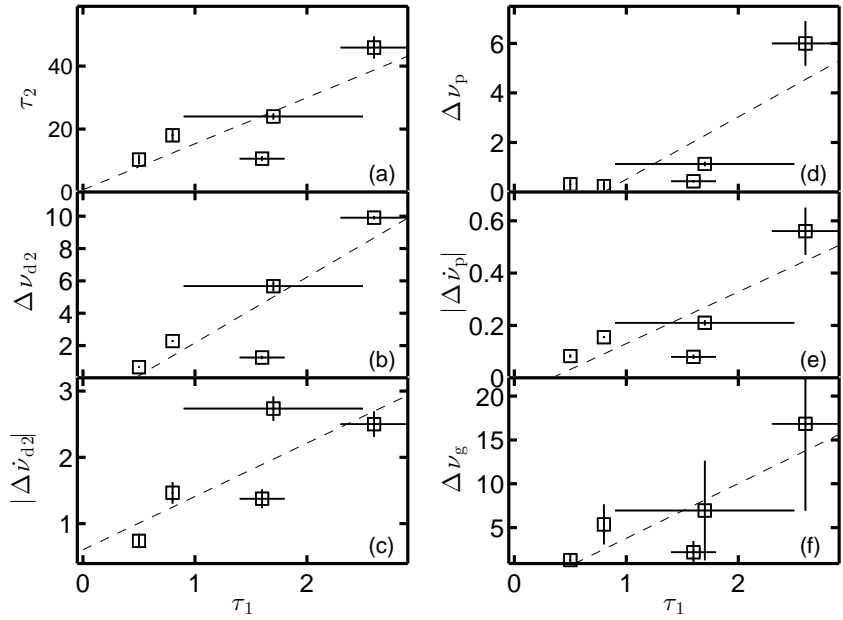}\caption{
The correlations between the parameters of G1--5 with delayed
spin-up processes. The dashed lines in panels (a)--(f) are the linear fitting
results for the correlations. The unit of $|\Delta\nu_{\rm d1}|$, $\Delta\nu_{\rm d2}$,  $\Delta\nu_{\rm p}$ and  $\Delta\nu_{\rm g}$
is $\mu$Hz. The unit of $|\Delta\dot\nu_{\rm d2}|$ and $|\Delta\dot\nu_{\rm p}|$
is pHz\,s$^{-1}$. $\tau_{1}$ and $\tau_{2}$ are in units of day.
\label{fig5}}
\end{center}
\end{figure}

Figure \ref{fig3} shows the relation between $\Delta\nu$ and
$|\Delta\dot\nu|$ for the Crab pulsar \citep{Espinoza2011}.
The Crab pulsar, PSR J0537$-$6910 and the Vela
pulsar have relatively large glitches as characterized by both
the large $\Delta\nu$ and $|\Delta\dot\nu|$ values. However, the glitch properties
of the Crab pulsar are very different from those of PSR J0537$-$6910 and the
Vela pulsar. $\Delta\nu$ and $|\Delta\dot\nu|$ of Crab's glitches have a strong positive
correlation \citep{Lyne2015}, in contrast to the other two pulsars without such
correlation \citep{Espinoza2011,Antonopoulou2018}. Five spin-up events are found for the Crab pulsar;
however, no similar spin-up phenomenon has been reported for either
PSR J0537--6910 or the Vela pulsar. The glitch around MJD 57734 from the
Vela pulsar is observed with the rising time scale less than 12.6\,s
\citep{Ashton et al.(2019)} and does not show
any evidence for delayed spin-up process.
Given the different ages of these three pulsars, we speculate that the states
of their crust sand interiors are different, and so the conditions of the physical
processes involved in glitches are different for pulsars with different ages.

In summary, in this work we have studied the glitches of the
Crab pulsar, which have delayed spin-up processes.
First, in addition to the three glitches occurred in 1989, 1996 and 2017,
we also found that second and fourth largest glitches of the Crab pulsar detected
in 2004 and 2011 have delayed spin-up processes, with the
second and fourth largest glitches detected in 2004 and 2011
are analyzed and these two glitches are found also with delayed spin-up processes of
$\tau_{1}=1.7\pm0.8$\,days and $\tau_{1}=1.6\pm0.4$\,days, respectively.
Using observations from {\sl Insight-HXMT}, Radio Telescopes in Xinjiang
and Kunming China and {\sl Fermi}, we studied the largest glitch in 2017
and obtained similar results with \cite{Shaw2018} and \cite{Zhang2018_2},
and further constrained its rising time to less than 0.48\,hour.

We obtained the correlations among the parameters of the delayed
spin-up processes and the parameters of the exponential decay processes:
the amplitudes of the delayed spin-up ($|\Delta{\nu}_{\rm d1}|$) and the
recovery process ($\Delta{\nu}_{\rm d2}$), their respective time scales ($\tau_{1}$, $\tau_{2}$),
and permanent changes of spin frequency ($\Delta{\nu}_{\rm p}$) have
strong positive correlations, while the rest parameters do not show any correlation
with each other. From the positive correlations, we suggest that further analysis
of the existing data for smaller glitches are needed to search for any evidence of
delayed spin-up processes with possibly shorter spin-up time scales, and more
high cadence observations of the Crab pulsar in the future are also critical in
understanding delayed spin-up processes and the interior structure of neutron stars.

\acknowledgments{This work is supported by the
National Key R\&D Program of China (2016YFA0400800)
and the National Natural Science Foundation of China under
grants U1838201, U1838202, U1938109, U1838104 and U1838101.
This work made use of the data from the HXMT mission,
a project funded by China National Space Administration (CNSA)
and the Chinese Academy of Sciences (CAS).
This work also made use of the radio observations from
Yunnan Observatory. The Nanshan 25\,m radio
telescope is jointly operated and administrated by Xinjiang Astronomical
Observatory and Center for Astronomical Mega-Science, Chinese Academy of Sciences.
We acknowledge the use of the public data from the {\sl RXTE}, {\sl INTEGRAL}
and {\sl Fermi} data archive.}

\clearpage

\begin{table}
\caption{Parameters of the Crab pulsar for G1--G5.}
\scriptsize{} \label{table_timing_para}
\begin{center}
\begin{tabular}{l l l l l l l}
\hline\hline
Parameters                                        & G1$^{(a)}$   &    G2$^{(a)}$     &G3      &         G4   &  G5           \\
\hline
Epoch(MJD)                                      &     --     &   --    &  53067                 &  55867            &   58038\\
$\nu$(Hz)                                         &     --     &    --   &  29.796943484(8)  &  $29.706916048(2)$ & $29.6375626144(3)$ \\
$\dot\nu$($10^{-10}$\,Hz\,s$^{-1}$)&    --   &   --        & -3.73308(5)        & $-3.70743(3)$  & $-3.686433(3)$ \\
$\ddot{\nu}$($10^{-20}$\,Hz\,s$^{-2}$)&   --   &  --      &   1.3(2)             & $0.96(17)$  & $0.81(3)$\\
\hline
Glitch epoch (MJD)                             & 47767.4       &  50259.93   & 53067.0780$^{(b)}$ & 55875.67(1)  &58064.548(2)\\
$ \Delta\nu_{p}$ ($\mu$Hz)               &   $2.38(2)$    &  $0.31(3)$   & $0.93(2)$       &  $0.49(3)$  & $5.7(9)$ \\
$ \Delta\dot\nu_{p}$ (pHz\,s$^{-1}$)  &    $-0.155(2)$ & $-0.083(6)$ & $-0.19(1)$     & $-0.09(1)$  & $-0.483(8)$ \\
$ \Delta\ddot\nu_{p}$ (10$^{-20}$Hz\,s$^{-2}$)&   -- &  $0.09(6)$   & --                     & --                       &  -- \\
$\Delta\nu_{d1}$ ($\mu$Hz)             &    -0.7             &    -0.31        &   $-0.35(5)$    & $-0.43(5)$  & $-1.23(1)$\\
$\tau_{1}$ (\,day)                              &    0.8              &   0.5             &   $1.7(8)$        &  $1.6(4)$    & $2.56(4)$\\
$\Delta\nu_{d2}$ ($\mu$Hz)             &     2.28           &    0.66          &  $5.67(4)$       &  $1.26(3)$  & $9.91(9)$\\
$\tau_{2}$ (\,day)                              &     18              &    10.3          &  $24(1)$          &   $10.6(3)$ & $45.9(3)$\\
Residuals ($\mu$s)                           &     --              &   --                &    214       &  311  &  113  \\
$\chi^{2}/{\rm d.o.f}$ (d.o.f)                &     --              &   --                &    0.99(52)       &  1.17(873)  &  1.35(1269)  \\
\hline
\end{tabular}
\end{center}
(a) The parameters are obtained form \cite{Wong2001}.\\
(b) Glitch epoch is adopted from http://www.atnf.csiro.au/people/pulsar/psrcat/glitchTbl.html\\
The confidence interval is 68.3\%.
\end{table}

\begin{table}
\caption{Pearson correlations between the parameters for G1--G5.}
\scriptsize{} \label{corre_para0}
\begin{center}
\begin{tabular}{l l l l l l l l l l l l}
\hline\hline
  ${\rm R}$ & $\tau_{1}$  &  $|\Delta\nu_{d1}|$ &$\Delta{\dot\nu}_{\rm d1}$& $\tau_{2}$ & $\Delta\nu_{d2}$  & $|\Delta{\dot\nu}_{\rm d2}|$ &  $ \Delta\nu_{p}$  &  $ \Delta\dot\nu_{p}$   & $ \Delta\nu_{\rm g}$  &$\hat{\rm Q}$ \\
\hline
$\tau_{1}$                           & 1 & 0.67 & -0.55 & 0.83 & 0.87 & 0.80 & 0.84 & -0.82 & 0.83 & -0.45 \\
$|\Delta\nu_{d1}|$               &  & 1 & 0.24 & 0.87 & 0.78 & 0.44 & 0.88 & -0.90 & 0.90 & -0.42 \\
$\Delta{\dot\nu}_{\rm d1}$ &  &  & 1 & -0.10& -0.25 & -0.50 & -0.14 & 0.10 & -0.10 & 0.22 \\
$\tau_{2}$                          &  &  &  & 1 & 0.98 & 0.76 & 0.95 & -0.99 & 0.99 & -0.43 \\
$|\Delta\nu_{d2}|$                &  &  &  &  & 1 & 0.86 & 0.92 & -0.96 & 0.97 & -0.4\\
$\Delta{\dot\nu}_{\rm d2}$ &  &  &  &  &  & 1 & 0.60 & -0.69 & 0.74 & 0.02 \\
$ \Delta\nu_{p}$                &  &  &  &  &  &  & 1 & -0.98 & 0.95 & -0.67 \\
$ \Delta\dot\nu_{p}$          &  &  &  &  &  &  &  & 1 & -0.99 & 0.52 \\
$ \Delta\nu_{\rm g}$          &  &  &  &  &  &  &  &  & 1 & -0.42 \\
$\hat{\rm Q}$                    &  &  &  &  &  &  &  &  &  & 1\\
\hline
\end{tabular}
\end{center}
\end{table}

\begin{table}
\caption{Spearman correlations between the parameters for G1--G5.}
\scriptsize{} \label{corre_para1}
\begin{center}
\begin{tabular}{l l l l l l l l l l l l}
\hline\hline
 ${\rm \rho}$ & $\tau_{1}$  &  $|\Delta\nu_{d1}|$ &$\Delta{\dot\nu}_{\rm d1}$& $\tau_{2}$ & $\Delta\nu_{d2}$  & $|\Delta{\dot\nu}_{\rm d2}|$ &  $ \Delta\nu_{p}$  &  $ \Delta\dot\nu_{p}$   & $ \Delta\nu_{\rm g}$  &$\hat{\rm Q}$ \\
\hline
$\tau_{1}$                          & 1 & 0.6 & -0.6 & 0.9 & 0.9 & 0.8 & 0.9 & -0.7 & 0.9 & -0.3 \\
$|\Delta\nu_{d1}|$               &   & 1 & 0.2 & 0.7 & 0.7 & 0.4 & 0.3 & -0.5 & 0.7 & -0.1 \\
$\Delta{\dot\nu}_{\rm d1}$ &    &    & 1 & -0.3 & -0.3 & -0.5 & -0.7 & 0.1 & -0.3 & 0.1 \\
$\tau_{2}$                          &    &    &    & 1 & 1.0 & 0.9 & 0.7 & -0.9 & 1.0 & -0.1 \\
$|\Delta\nu_{d2}|$              &    &    &    &   & 1 & 0.9 & 0.7 & -0.9 & 1.0 & -0.1\\
$\Delta{\dot\nu}_{\rm d2}$ &    &    &    &    &   & 1 & 0.6 & -0.8 & 0.9 & 0.2 \\
$ \Delta\nu_{p}$                &    &    &    &    &    &    & 1 & -0.6 & 0.7 & -0.6 \\
$ \Delta\dot\nu_{p}$          &    &    &    &    &    &    &    & 1 & -0.9 & 0.2 \\
$ \Delta\nu_{\rm g}$          &    &    &    &    &    &    &    &    & 1 & -0.1 \\
$\hat{\rm Q}$                    &    &    &    &    &    &     &    &    &   & 1\\
\hline
\end{tabular}
\end{center}
\end{table}

\end{document}